\documentclass[aps,pre,eqsecnum,preprint,tightenlines,nofootinbib,showpacs]{revtex4-1}
\usepackage{graphicx,amsmath,latexsym}

\def\bea{\begin{eqnarray}}
\def\eea{\end{eqnarray}}
\def\be{\begin{equation}}
\def\ee{\end{equation}}

\newcommand{\Pminus}{{\cal P}^-}

\begin{document}

\title{Symmetric multivariate polynomials \\
as a basis for three-boson light-front wave functions}

\author{Sophia S. Chabysheva}
\author{Blair Elliott}
\author{John R. Hiller}
\affiliation{Department of Physics \\
University of Minnesota-Duluth \\
Duluth, Minnesota 55812}

\date{\today}

\begin{abstract}
We develop a polynomial basis to be used in numerical calculations of light-front Fock-space
wave functions.  Such wave functions typically depend on longitudinal momentum fractions that
sum to unity.  For three particles, this constraint limits the two remaining independent
momentum fractions to a triangle, for which the three momentum fractions act as barycentric
coordinates. For three identical bosons, the wave function must be symmetric with respect
to all three momentum fractions.  Therefore, as a basis, we construct polynomials in two
variables on a triangle that are symmetric with respect to the interchange of any two
barycentric coordinates.  We find that, through the fifth order, the polynomial is unique
at each order, and, in general, these polynomials can be 
constructed from products of powers of the second and third-order
polynomials.  The use of such a basis is illustrated in a calculation of a 
light-front wave function in two-dimensional $\phi^4$ theory; the polynomial basis
performs much better than the plane-wave basis used in discrete light-cone quantization.
\end{abstract}
%%%%%%%%%%%%%%%%%%%%%%%%%%%%%%%%%%%%%%%%%%%%%%%%%%%%%
%02.60.Cb 	Numerical simulation; solution of equations 
%02.60.Nm 	Integral and integrodifferential equations
%11.10.Ef 	Lagrangian and Hamiltonian approach 
% 10. THE PHYSICS OF ELEMENTARY PARTICLES AND FIELDS
% 11.10.-z Field theory (for gauge field theories, see 11.15)
% 11.10.Ef Lagrangian and Hamiltonian approach
% 11.10.Gh Renormalization
% 11.15.-q Gauge field theories
% 11.15.Tk Other nonperturbative techniques
% 12. Specific theories and interaction models; particle systematics
% 12.38.-t Quantum chromodynamics see also 24.85 Quarks, gluons,
% and QCD in nuclei and nuclear processes
% 12.38.Lg Other nonperturbative calculations
%%%%%%%%%%%%%%%%%%%%%%%%%%%%%%%%%%%%%%%%%%%%%%%%%%%%%%%

%
\pacs{11.15.Tk, 11.10.Ef, 02.60.Nm
}

\maketitle

%%%%%%%%%%%%%%%%%%%%%%
\section{Introduction}
\label{sec:introduction}
%%%%%%%%%%%%%%%%%%%%%%

Light-front quantization~\cite{Dirac,DLCQreview} is a natural choice
for the nonperturbative solution of a quantum field theory.  The
eigenstates are built as expansions in terms of Fock states, states
of definite particle number and definite momentum, where the coefficients
are boost-invariant wave functions.  The vacuum state is simply the
Fock vacuum, thereby giving the wave functions a standard, quantum mechanical
interpretation.

The light-front time coordinate is chosen to
be $x^+\equiv t+z/c$, and the corresponding light-front spatial 
coordinate is $x^-\equiv t-z/c$; the other spatial coordinates are
unchanged.  The conjugate light-front energy is $p^-=E-cp_z$, and
the light-front longitudinal momentum is $p^+=E/c+p_z$.
A boost-invariant momentum fraction $x_i=p_i^+/P^+$ is defined
for the ith particle with momentum $p_i^+$ in a system with 
total momentum $P^+$.  Because the light-front longitudinal
momentum is always positive, these momentum fractions are between
zero and one.  Also, momentum conservation dictates that they sum to one.

In the three-particle case, the three momentum fractions correspond
to the barycentric coordinates of a triangle.  Any two can be
treated as the independent variables.  For a wave function
that describes three identical bosons, there must be symmetry
under the interchange of any two of the three coordinates, not just
symmetry under the interchange of
the two chosen as independent.  Any set of basis functions to be
used in numerical approximations of such a wave function should
share this symmetry.  However, the usual treatment of two-variable 
polynomials on a triangle is limited to consideration of symmetry
with respect to only the two independent variables~\cite{sympoly,multpoly}.
Here we consider the full-symmetry constraint.

We find that full symmetry among all three barycentric coordinates
dramatically reduces the number of polynomials at any given
order.  For the lowest orders, there is only one; at the sixth
order, there are two.  In general,
for polynomials of order $N$, the number of linearly
independent polynomials is the number of combinations of two
nonnegative integers $n$ and $m$ such that $N=2n+3m$.  These
polynomials can be chosen to be products of $n$ factors of
the second-order polynomial and $m$ factors of the third-order
polynomial.  They are not orthonormal, but given such a set
of polynomials one can, of course, systematically generate
an orthonormal set.

As a test of the utility of these polynomials, we consider
a problem in two-dimensional $\phi^4$ theory where the mass of
the eigenstate is shifted by coupling between the one-boson
sector and the three-boson sector.  The results obtained
are quite encouraging.  For comparison we also consider
discrete light-cone quantization (DLCQ)~\cite{PauliBrodsky,DLCQreview}
which uses a periodic plane-wave basis and therefore 
quadratures in momentum space that use equally spaced 
points.  The DLCQ results would require extrapolation to
obtain an accurate answer, whereas the symmetric-polynomial basis
immediately converges.

The content of the remainder of the paper is as follows.  In Sec.~\ref{sec:polynomials},
we specify the construction of the fully symmetric polynomials.  The first
subsection describes the lowest order cases, where a first-order polynomial
is found to be absent and the second and third-order polynomials are found
to be unique.  The second subsection gives the analysis at any finite order,
with details of a proof left to an Appendix.
The illustration of the use of these polynomials, as a basis for
the three-boson wave function in $\phi^4$ theory, is presented in
Sec.~\ref{sec:illustration}.  A brief summary is given in Sec.~\ref{sec:summary}.

%%%%%%%%%%%%%%%%%%%%%%
\section{Fully symmetric polynomials}
\label{sec:polynomials}
%%%%%%%%%%%%%%%%%%%%%%

%%%%%%%%%%%%%%%%%%%%%%
\subsection{Lowest orders}
\label{sec:loworders}
%%%%%%%%%%%%%%%%%%%%%%

We consider polynomials in Cartesian coordinates $x$ and $y$,
on the triangle defined by 
\be
0\leq x \leq 1, \;\;0\leq y \leq 1, \;\; 0\leq 1-x-y \leq 1,
\ee
that are fully symmetric with respect to interchange of
the coordinates $x$, $y$, and $z=1-x-y$.
These can be viewed as the restriction of three-variable
polynomials on the unit cube to the plane $x+y+z=1$.
The construction of the fully symmetric three-variable
polynomials on the cube is trivial; at order $N$, the
possible polynomials are linear combinations of the form
\be
x^i y^j z^k+ x^j y^k z^i +x^k y^i z^j +x^j y^i z^k + x^i y^k z^j+x^k y^j z^i,
\ee
with $i$, $j$, and $k$ nonnegative integers such that $N=i+j+k$.
The linearly independent polynomials would correspond to some
particular ordering of these indices, such as $i\leq j\leq k$.
For $N=0$ or 1 there is only one polynomial, but for $N\geq2$ there
are several.

The restriction to the plane defined by $x+y+z=1$ is, however,
a severe constraint.  As we will see, the fully symmetric 
two-variable polynomials are unique up through $N=5$.
For $N=1$, the constraint eliminates the only candidate; the
restriction from the cube to the plane makes $x+y+z$ just
a constant.  For $N=2$, we have two candidates
\be
x^2+y^2+z^2 \;\;\mbox{and}\;\; xy+xz+yz.
\ee
Substitution of $z=1-x-y$ quickly shows that they are equivalent
up to terms of order less than two.  Similarly, for $N=3$, the
three candidates
\be
x^3+y^3+z^3, \;\; x^2y+x^2z+xy^2+xz^2+y^2z+yz^2, \;\; \mbox{and} \;\;xyz
\ee
reduce to equivalent polynomials, up to terms of order less than three,
upon substitution of $z=1-x-y$.  Equivalence does not exclude the
possibility that the polynomials will differ by fully symmetric
polynomials of lower order.  The terms of order three are the same,
and the polynomials differ by at most symmetric polynomials of
lower order.

To proceed in this fashion to higher orders is, of course,
possible but tedious.  Instead we develop a direct analysis
of the possible two-variable polynomials and the symmetry
constraints, as described in the next subsection.

%%%%%%%%%%%%%%%%%%%%%%
\subsection{General analysis}
\label{sec:general}
%%%%%%%%%%%%%%%%%%%%%%

In order to avoid complications due to lower-order contributions,
we first change variables from $x,y,z$ to $u,v,w$ defined by
\be \label{eq:uvw}
u=x-1/3,\;\; v=y-1/3, \;\; w=z-1/3=-(u+v).
\ee
Any polynomial $P$ on the triangle, for which each term is of order $N$,
can be written in the form 
\be
P(u,v)=\sum_{n=0}^N c_n u^n v^{N-n},
\ee
and, unlike replacement of $x$ or $y$ by $z=1-x-y$, powers of $w=-(u+v)$
that appear in replacements of $u$ or $v$ do not introduce
lower-order contributions.

Symmetry with respect to just $u$ and $v$ restricts the
coefficients to be such that $c_n=c_{N-n}$.  If symmetry
with respect to $v\rightarrow w=-(u+v)$ is imposed,
the coefficients must satisfy the constraint
\be \label{eq:u+v}
\sum_{n=0}^N c_n u^{N-n} v^n = \sum_{n=0,N} c_n u^{N-n} (-1)^n (u+v)^n
=\sum_{n=0}^N c_n u^{N-n} (-1)^n\sum_{m=0}^n \left(\begin{array}{c} n \\ m\end{array}\right) u^{n-m}v^m.
\ee
These are sufficient to guarantee that the resulting polynomial has all the
desired symmetries.

The symmetry conditions can be reduced to a linear system for the
coefficients.  With a change in the order of the sums on the right
of (\ref{eq:u+v}) and an interchange of the summation indices $m$ and $n$, we find
\be
\sum_{n=0}^Nc_n u^{N-n} v^n 
=\sum_{n=0}^N \sum_{m=n}^N (-1)^m \left(\begin{array}{c} m \\ n\end{array}\right)c_m u^{N-n} v^n.
\ee
Therefore, the coefficients must satisfy the linear system
\be
c_n=c_{N-n}, \;\; \sum_{m=n}^N (-1)^m \left(\begin{array}{c} m \\ n\end{array}\right)c_m=c_n.
\ee
This system may at first seem to be overdetermined, but instead it is typically underdetermined.
A solution exists for any $N$ other than $N=1$.
For $N=2,3,4$, and 5, there is one linearly independent solution;
and, for $N\geq6$, there can be two or more linearly independent solutions.

For example, with $N=6$ the system can be expressed in matrix form as
\be
\left(
\begin{array}{ccccccc}
 0 & -1 & 1 & -1 & 1 & -1 & 1 \\
 0 & -2 & 2 & -3 & 4 & -5 & 6 \\
 0 & 0 & 0 & -3 & 6 & -10 & 15 \\
 0 & 0 & 0 & -2 & 4 & -10 & 20 \\
 0 & 0 & 0 & 0 & 0 & -5 & 15 \\
 0 & 0 & 0 & 0 & 0 & -2 & 6 \\
 0 & 0 & 0 & 0 & 0 & 0 & 0
\end{array}
\right)
\left(\begin{array}{c} c_0 \\ c_1 \\ c_2 \\ c_3 \\ c_2 \\ c_1 \\ c_0 \end{array}\right)=
\left(\begin{array}{c} 0\\ 0 \\ 0 \\ 0 \\ 0 \\ 0 \\ 0 \end{array}\right).
\ee
The determinant is obviously zero, as is the case for any $N$, allowing nontrivial solutions.
The system reduces to two equations
\be
3c_0-c_1=0,\;\; 5 c_0 - 2 c_2 + c_3=0
\ee
for the four unknowns, leaving two linearly independent solutions, such as
\be
u^6+3 u^5 v +5 u^3v^3+3 u v^5 +v^6\;\;  \mbox{and} \;\;
u^4 v^2+2 u^3v^3+u^2 v^4.
\ee

For any value of $N$, one finds that the number of independent solutions
is always the number of ways that $N$ can be written as $2n+3m$ for nonnegative
integers $n$ and $m$.  A proof of this conjecture for arbitrary $N$
is given in the Appendix.  Thus, in each of these cases, a fully symmetric 
polynomial can be chosen to be the product of $n$ copies of the second-order
polynomial and $m$ copies of the third-order polynomial, or a linear 
combination of such polynomials.  Returning to the original Cartesian
coordinates, we take these two base polynomials to be
\be \label{eq:basepolys}
C_2(x,y)=x^2+y^2+(1-x-y)^2\;\; \mbox{and} \;\;
C_3(x,y)=xy(1-x-y).
\ee
We then have that all fully symmetric polynomials can be constructed
from linear combinations of the products
\be \label{eq:Cnm}
C_{nm}(x,y)=C_2^n(x,y) C_3^m(x,y).
\ee

These do not form an orthonormal set.  To construct such a set, we
apply the Gramm--Schmidt process, relative to the inner product
\be
\int_0^1 dx \int_0^{1-x} dy P_n^{(i)}(x,y) P_m^{(j)}(x,y) =\delta_{nm}\delta_{ij},
\ee
where $P_n^{(i)}$ is the ith polynomial of order $n$.  
The first few polynomials are
\bea \label{eq:firstfew}
P_0&=&\sqrt{2}, \\
P_1&=&0, \nonumber \\
P_2&=&\sqrt{30} \left[4  x^2+4  y x-4  x+4  y^2-4  y+1\right],
  \nonumber \\
P_3&=&\sqrt{3}\left[-140  y x^2+20  x^2-140  y^2 x+160  y x
      -20  x+20  y^2-20  y+8/3\right],
    \nonumber \\
P_4&=&\sqrt{42} \left[60  x^4+120  y x^3-120  x^3+180  y^2 x^2
    -200  y x^2+80  x^2+120    y^3 x-200  y^2 x \right. \nonumber \\
    && \left. \rule{0.5in}{0mm} +100  y x-20  x  +60  y^4-120  y^3+80  y^2-20 y+5/3 \right],
    \nonumber \\
P_5&=&\sqrt{6}\left[-2310  y x^4+210  x^4-4620  y^2 x^3+5040  y x^3
       -420  x^3-4620  y^3 x^2+7560
    y^2 x^2\right. \nonumber \\
    && \left. \rule{0.5in}{0mm} -3780  y x^2+280  x^2-2310  y^4 x
         +5040  y^3 x-3780  y^2 x+1120
    y x-70  x\right. \nonumber \\
    && \left. \rule{0.5in}{0mm} +210  y^4-420  y^3+280  y^2-70  y+4 \right],
    \nonumber \\
P_6^{(1)}&=&\sqrt{\frac{10}{11863}}\left[240240  x^6+720720   y x^5
                  -720720   x^5+1441440
     y^2 x^4-1829520   y x^4\right. \nonumber \\
    && \left. \rule{0.5in}{0mm} +826980   x^4+1681680
     y^3 x^3-2938320   y^2 x^3+1709400   y x^3-452760
     x^3\right. \nonumber \\
    && \left. \rule{0.5in}{0mm} +1441440   y^4 x^2-2938320   y^3 x^2+2203740
     y^2 x^2-733320   y x^2+120204   x^2\right. \nonumber \\
    && \left. \rule{0.5in}{0mm} +720720
     y^5 x-1829520   y^4 x+1709400   y^3 x-733320
     y^2 x+146664   y x\right. \nonumber \\
    && \left. \rule{0.5in}{0mm} -13944   x+240240
     y^6-720720   y^5+826980   y^4-452760
     y^3+120204   y^2\right. \nonumber \\
    && \left. \rule{0.5in}{0mm} -13944   y+581 \right],
    \nonumber \\
P_6^{(2)}&=&\sqrt{\frac{143}{11863}}\left[ -16436  x^6-49308   y x^5+49308   x^5+399630
     y^2 x^4-28140   y x^4-50190   x^4\right. \nonumber \\
    && \left. \rule{0.5in}{0mm} +881440
     y^3 x^3-1102080   y^2 x^3+202440   y x^3+18200
     x^3+399630   y^4 x^2\right. \nonumber \\
    && \left. \rule{0.5in}{0mm} -1102080   y^3 x^2+826560
     y^2 x^2-155400   y x^2-210   x^2-49308
     y^5 x-28140   y^4 x\right. \nonumber \\
    && \left. \rule{0.5in}{0mm} +202440   y^3 x-155400
     y^2 x+31080   y x-672   x-16436
     y^6+49308   y^5\right. \nonumber \\
    && \left. \rule{0.5in}{0mm} -50190   y^4+18200
     y^3-210   y^2-672   y+28 \right].
    \nonumber
\eea
If there is only one polynomial at a particular order, the $i$ index is dropped.

%%%%%%%%%%%%%%%%%%%%%%
\section{Illustration}
\label{sec:illustration}
%%%%%%%%%%%%%%%%%%%%%%

As a sample application, we consider the integral equation for
the three-boson wave function in two-dimensional $\phi^4$ theory.
This equation is obtained from the fundamental Hamiltonian
eigenvalue problem on the light front~\cite{DLCQreview},
\be
\Pminus|\psi(P^+)\rangle=\frac{M^2}{P^+}|\psi(P^+)\rangle \;\;
\mbox{and} \;\; 
{\cal P}^+|\psi(P^+)\rangle=P^+|\psi(P^+)\rangle.
\ee
The second equation is automatically satisfied by expanding
the eigenstate in Fock states $|p_i^+;P^+,n\rangle$ of $n$ bosons
with momentum $p_i^+$ such that $\sum_i p_i^+=P^+$:
\be
|\psi(P^+)\rangle=\sum_n (P^+)^{(n-1)/2}
    \int \left(\prod_{i=1}^{n-1} dx_i\right) \psi_n(x_1,...,x_n) |x_iP^+;P^+,n\rangle.
\ee
Here $\psi_n$ is the $n$-boson wave function, and the factor $(P^+)^{(n-1)/2}$
is explicit in order that $\psi_n$ be independent of $P^+$.

The light-front Hamiltonian for $\phi^4$ theory is
\bea \label{eq:Pminus}
\Pminus&=&\int dp^+ \frac{\mu^2}{p^+} a^\dagger(p^+)a(p^+)  \\
  && +\frac{\lambda}{6}\int \frac{dp_1^+dp_2^+dp_3^+}
                              {4\pi \sqrt{p_1^+p_2^+p_3^+(p_1^++p_2^++p_3^+)}} \nonumber \\
  &&\rule{1in}{0mm} 
    \times \left[a^\dagger(p_1^++p_2^++p_3^+)a(p_1^+)a(p_2^+)a(p_3^+)\right. \nonumber \\
  && \rule{1.25in}{0mm} \left. 
    +a^\dagger(p_1^+)a^\dagger(p_2^+)a^\dagger(p_3^+)a(p_1^++p_2^++p_3^+)\right]  \nonumber \\
 && +\frac{\lambda}{4}\int\frac{dp_1^+ dp_2^+}{4\pi\sqrt{p_1^+p_2^+}}
       \int\frac{dp_1^{\prime +}dp_2^{\prime +}}{\sqrt{p_1^{\prime +} p_2^{\prime +}}} 
       \delta(p_1^+ + p_2^+-p_1^{\prime +}-p_2^{\prime +}) \nonumber \\
 && \rule{2in}{0mm} \times a^\dagger(p_1^+) a^\dagger(p_2^+) a(p_1^{\prime +}) a(p_2^{\prime +}) .
   \nonumber
\eea
The mass of the constituent bosons is $\mu$, and $\lambda$ is the coupling constant.
The operator $a^\dagger(p^+)$ creates a boson with momentum $p^+$; it obeys the commutation
relation
\be
[a(p^+),a^\dagger(p^{\prime +}]=\delta(p^+-p^{\prime +})
\ee
and builds the Fock states from the Fock vacuum $|0\rangle$ in the form
\be
|x_iP^+;P^+,n\rangle=\frac{1}{\sqrt{n!}}\prod_{i=1}^n a^\dagger(x_iP^+)|0\rangle.
\ee
The terms of the light-front Hamiltonian are such that $\Pminus$ changes particle
number not at all or by two; therefore, the number of constituents in a
contribution to the eigenstate is always either odd or even.

We consider the odd case, and, to have a finite eigenvalue problem, we truncate
the Fock-state expansion at three bosons.  We also simplify to a problem
with an exact solution by dropping from the Hamiltonian the two-body scattering
term, the last term in (\ref{eq:Pminus}).  The action of the light-front
Hamiltonian then yields the following coupled system of integral equations:
\bea
M^2\psi_1&=&\mu^2\psi_1
  +\frac{\lambda}{\sqrt{6}}\int\frac{dx_1 dx_2}{4\pi\sqrt{x_1 x_2 x_3}}\psi_3(x_1,x_2,x_3), \\
M^2\psi_3&=&\mu^2\left(\frac{1}{x_1}+\frac{1}{x_2}+\frac{1}{x_3}\right)\psi_3
  +\frac{\lambda}{\sqrt{6}}\frac{\psi_1}{4\pi\sqrt{x_1x_2x_3}}.
\eea
It is understood that $x_3=1-x_1-x_2$.

To create a single integral equation for $\psi_3$, we use the first equation
to eliminate $\psi_1$ from the second, leaving
\be
M^2\psi_3=\mu^2\left(\frac{1}{x_1}+\frac{1}{x_2}+\frac{1}{x_3}\right)\psi_3
  -\frac{\lambda^2}{6(4\pi)^2}\frac{1}{\mu^2-M^2}\frac{1}{\sqrt{x_1x_2x_3}}
     \int \frac{dx'_1 dx'_2}{\sqrt{x'_1 x'_2 x'_3}}\psi_3(x'_1,x'_2,x'_3).
\ee
This is no longer a simple eigenvalue problem for $M^2$, but it can be
rearranged into an eigenvalue problem for the reciprocal of a 
dimensionless coupling, defined as
\be
\xi=6(1-M^2/\mu^2)\left(\frac{4\pi\mu^2}{\lambda}\right)^2.
\ee
The rearrangement yields
\be
\xi\psi_3=\left[\frac{1}{x_1}+\frac{1}{x_2}+\frac{1}{x_3}-\frac{M^2}{\mu^2}\right]^{-1}
\frac{1}{\sqrt{x_1x_2x_3}}
     \int \frac{dx'_1 dx'_2}{\sqrt{x'_1 x'_2 x'_3}}\psi_3(x'_1,x'_2,x'_3).
\ee
To symmetrize the kernel of this equation, we replace $\psi_3$ by
\be \label{eq:f3}
\psi_3(x_1,x_2,x_3)=\left[\frac{1}{x_1}+\frac{1}{x_2}+\frac{1}{x_3}-\frac{M^2}{\mu^2}\right]^{-1/2}f_3(x_1,x_2,x_3)
\ee
and obtain
\bea
\xi f_3&=&\frac{1}{\sqrt{x_1x_2x_3}}
      \left[\frac{1}{x_1}+\frac{1}{x_2}+\frac{1}{x_3}-\frac{M^2}{\mu^2}\right]^{-1/2} \\
&& \rule{0.5in}{0mm} \times 
     \int  \frac{dx'_1 dx'_2}{\sqrt{x'_1 x'_2 x'_3}}
     \left[\frac{1}{x'_1}+\frac{1}{x'_2}+\frac{1}{x'_3}-\frac{M^2}{\mu^2}\right]^{-1/2}
     f_3(x'_1,x'_2,x'_3). \nonumber
\eea
This rearrangement also accomplishes an important step toward the use of a polynomial
expansion.  The leading small-$x_i$ behavior of $\psi_3$ is $\sqrt{x_i}$, and, as
can be seen from the structure of the pre-factor in (\ref{eq:f3}), the leading behavior
of $f_3$ is just a constant.

Because the kernel factorizes, the equation can be solved analytically.  The
function $f_3$ must be of the form
\be
f_3(x_1,x_2,x_3)=\frac{A}{\sqrt{x_1x_2x_3}}
      \left[\frac{1}{x_1}+\frac{1}{x_2}+\frac{1}{x_3}-\frac{M^2}{\mu^2}\right]^{-1/2},
\ee
with a normalization $A$.  Substitution of this form into the equation
for $f_3$ yields the condition for the eigenvalue:
\be \label{eq:xi}
\xi=\int \frac{dx_1 dx_2}{x_1 x_2 x_3}
   \left[\frac{1}{x_1}+\frac{1}{x_2}+\frac{1}{x_3}-\frac{M^2}{\mu^2}\right]^{-1}.
\ee
A value can be computed when the ratio $M/\mu$ is specified.

To solve the equation for $f_3$ with the symmetric polynomial basis, we 
substitute the truncated expansion
\be
f_3=\sum_{n,i}^N a_{ni}P_n^{(i)}
\ee
and obtain a matrix eigenvalue problem for the coefficients
\be
\sum_{m,j}^N b_{ni} b_{mj} a_{mj} =\xi a_{ni},
\ee
with
\be
b_{ni}\equiv\int\frac{dx_1 dx_2}{\sqrt{x_1 x_2 x_3}}
  \frac{P_n^{(i)}(x_1,x_2,x_3)} {\sqrt{1/x_1+1/x_2+1/x_3-M^2/\mu^2}} .
\ee
The eigenvalue is then approximated by 
\be
\xi \simeq \sum_{n,i}^N b_{ni}^2.
\ee
A set of values for different $N$ is given in Table~\ref{tab:example} for $M^2=0.5\mu^2$.
The convergence to the exact value is quite rapid.  Similar behavior occurs for other
values of $M$.
%%%%%%%%%%%%%%%%%%%%%%%%%%%%%%%%%%%%%
\begin{table}[ht]
\caption{\label{tab:example} Sequence of eigenvalue approximations obtained with use of the fully
symmetric polynomials $P_N^{(i)}$ up to the eighth order for $M^2=\frac12\mu^2$.  Orders
six and eight appear twice, because there are two polynomials in each case; however, the
result changes little with the addition of the second polynomial.  These results are
to be compared with the exact value of $\xi=2.40335$.
}
\begin{center}
\begin{tabular}{c|cccccccccc}
\hline \hline
$N$ & 1 & 2 & 3 & 4 & 5 & 6 & 6 & 7 & 8 & 8 \\
\hline
$\xi$ & 2.25637 & 2.35351 & 2.36321 & 2.38048 & 2.38489 & 2.39040 & 2.39057 & 2.39273 & 2.39504 & 2.39525 \\
\hline \hline
\end{tabular}
\end{center}
\end{table}
%%%%%%%%%%%%%%%%%%%%%%%%%%%%%%%%%%%%%%%%%%%%%%%%%%%%%%%%%%%%%%

\begin{figure}[ht]
\vspace{0.2in}
\centerline{\includegraphics[width=10cm]{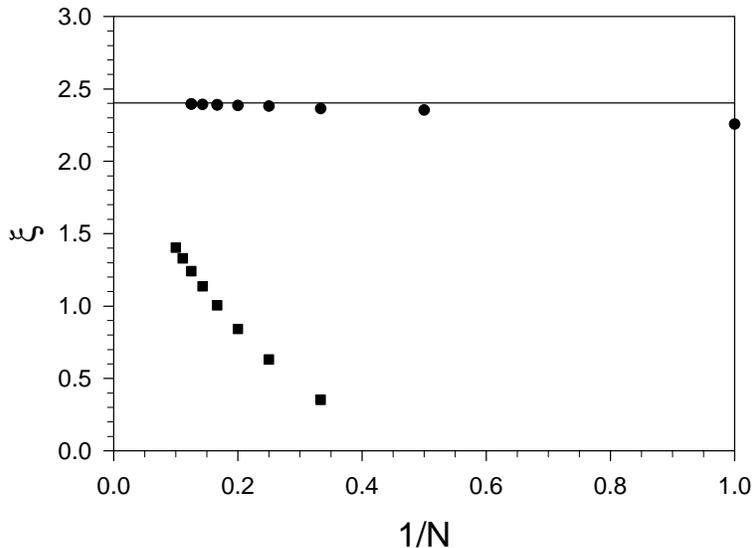}}
\caption{\label{fig:compare} Comparison of convergence rates for the
fully symmetric polynomial basis (filled circles) and DLCQ (filled squares).  
The dimensionless eigenvalue $\xi$ is plotted versus $1/N$, the reciprocal of the basis
order and of the DLCQ resolution, for the case where $M^2=0.5\mu^2$.  The 
horizontal line is at the exact value, $\xi=2.40335$.
}
\end{figure}

By way of comparison, we also consider the DLCQ approach.  In the present
circumstance, DLCQ yields a trapezoidal approximation to the integral in Eq.~(\ref{eq:xi}),
with the step sizes in $x_1$ and $x_2$ taken as $1/N$ for an integer
resolution $N$.  Points on the edge of the triangle, which correspond to zero-momentum
modes, are usually ignored.  The DLCQ approximation is then
\be
\xi\simeq\frac{1}{N^2}\sum_{i=1}^{N-2}\sum_{j=1}^{N-i-1} \frac{N^3}{i j (N-i-j)}
   \left[\frac{N}{i}+\frac{N}{j}+\frac{N}{N-i-j}-\frac{M^2}{\mu^2}\right]^{-1}.
\ee
Results for the two approximations are presented in Fig.~\ref{fig:compare}.
The symmetric polynomial approximation converges much faster.  The primary difficulty
for the DLCQ approximation is the integrable singularity at each corner of the
triangle.\footnote{To be fair, we should point out that DLCQ is used primarily
for many-body problems, where basis function expansions are difficult to implement,
and can be combined with an extrapolation procedure to obtain converged results.}
%
%%%%%%%%%%%%%%%%%%%%%%%%%%%%%%%%%%%%%
%%%%%%%%%%%%%%%%%%%%%%%%%%%%%%%%%%%%%%%%%%%%%%%%%%%%%%%%%%%%%%

%%%%%%%%%%%%%%%%%%%%%%
\section{Summary}
\label{sec:summary}
%%%%%%%%%%%%%%%%%%%%%%

We have constructed an orthonormal set of fully symmetric polynomials on a triangle
that can be used as a basis for three-boson longitudinal wave functions in field
theories quantized on the light front~\cite{Dirac,DLCQreview}.  At each order, the
number of polynomials is quite small, the limitation to symmetry under
the interchange of all three barycentric coordinates being a much stronger
constraint than just symmetry under interchange of the two independent
variables.  A list of the first six polynomials is given in Eq.~(\ref{eq:firstfew}).
In general, the polynomials are formed by first constructing a non-orthonormal
set according to Eq.~(\ref{eq:Cnm}), and then applying an orthogonalizing procedure,
such as the Gramm--Schmidt process.

As a sample application, we have considered a light-front Hamiltonian eigenvalue
problem in $\phi^4$ theory, limited to the coupling of one-boson and three-boson
Fock states.  The polynomial expansion for the wave function yields rapidly 
converging results, particularly in comparison with a DLCQ approximation,
as can be seen in Table~\ref{tab:example} and Fig.~\ref{fig:compare}.

The original motivation for these developments was to find an expansion 
applicable to the nonlinear equations of the light-front coupled-cluster
(LFCC) method~\cite{LFCC}.  In this method, there is no truncation of
Fock space, but approximations for the wave functions for higher Fock states
are determined from the wave functions of the lowest states by functions that satisfy
nonlinear integral equations.  In bosonic theories, these functions must have the full
symmetry, and any basis used should have this symmetry.  The sample application here
can be interpreted as a linearization of the $\phi^4$ LFCC equations.  Thus, we expect
the new polynomial basis to be of considerable utility.

\acknowledgments
This work was supported in part by the Department of Energy
through Contract No.\ DE-FG02-98ER41087.

\appendix

%%%%%%%%%%%%%%%%%%%%%%%%%%%%%%%%%%%%%%%%%
\section{Proof of the conjecture}  \label{sec:proof}
%%%%%%%%%%%%%%%%%%%%%%%%%%%%%%%%%%%%%%%%%%%

Here we give a proof that any fully symmetric polynomial on a
triangle can be expressed as a linear combination of products
of powers of two fundamental polynomials of order two and three.
We work in terms of the translated variables $u$, $v$, and $w$
defined in (\ref{eq:uvw}), so that the constraint of being on
the triangle is $u+v+w=0$.  The structure of the proof is first to
characterize unconstrained polynomials on the unit cube and then
to restrict these polynomials to the triangle.

Any symmetric polynomial built from mononials of order $N$ is
a linear combination of polynomials $\tilde{P}_{ijk}(u,v,w)$
defined by
\be
\tilde{P}_{ijk}(u,v,w)=u^i v^j w^k +\,\mbox{permutations},
\ee
with $i+j+k=N$ and $i\leq j\leq k$. Thus, the $\tilde{P}_{ijk}$
form a basis for symmetric three-variable polynomials with each
term of order $N$.  The size of this basis is
\be
S_N\equiv \sum_{i=0}^{[N/2]}\sum_{j=i}^{[(N-i)/2]} 1,
\ee
where $[x]$ means the integer part of $x$.  The limits on
the sums guarantee the order $i\leq j\leq k$, with $k=N-i-j$.

We can also build symmetric polynomials from linear
combinations of 
\be
\tilde{C}_{lnm}(u,v,w)=\tilde{C}_1^l(u,v,w)\tilde{C}_2^n(u,v,w)\tilde{C}_3^m(u,v,w),
\ee
where
\be
\tilde{C}_1=u+v+w,\;\;
\tilde{C}_2=uv+uw+vw,\;\;
\tilde{C}_3=uvw,
\ee
and $N=l+2n+3m$.  However, is this sufficient to generate all such polynomials?
The number of polynomials $\tilde{C}_{lmn}$ is
\be
\Xi_N\equiv \sum_{m=0}^{[N/3]}\sum_{n=0}^{[(N-3m)/2]} 1,
\ee
which counts the number of ways that the integers $l$, $n$, and $m$ can be assigned,
with $l=N-2n-3m$.  The substitutions $m=i$ and $n=j-i$ yield
\be
\Xi_N=\sum_{i=0}^{[N/3]}\sum_{j=i}^{[(N-i)/2-i]+i} 1
     =\sum_{i=0}^{[N/3]}\sum_{j=i}^{[(N-i)/2]} 1.
\ee
Therefore, $\Xi_N$ is equal to $S_N$, and the $\tilde{C}_{lnm}$ do form an
equivalent basis on the unit cube.

The projection onto the triangle $u+v+w=0$ eliminates $\tilde{C}_1$ and any 
basis polynomial $\tilde{C}_{lnm}$ with $l>0$.  Thus, the basis polynomials
on the triangle can be chosen as products of powers of second and third-order
polynomials.  The powers $n$ and $m$, respectively, include all possible
integers that satisfy $N=2n+3m$.  In terms of the Cartesian variables $x$
and $y$, we then have the basis set specified by (\ref{eq:basepolys}) and
(\ref{eq:Cnm}).

%%%%%%%%%%%%%%%%%%%%%%%%%%%%%%%%

\end{document}